\documentclass[%
reprint,
 groupedaddress,
 twocolumn,
 amsmath,amssymb,
 aps,
 prb,
]{revtex4-1}

\usepackage{graphicx}
\usepackage{dcolumn}
\usepackage{bm}
\usepackage{braket} 
\usepackage[colorlinks=true, linkcolor=cyan, urlcolor=blue]{hyperref}
\usepackage[export]{adjustbox} 
\usepackage{tabularx}

\usepackage[flushleft]{threeparttable}
\usepackage{tikz}
\usetikzlibrary{calc}

\begin{document}

\preprint{APS/123-QED}

\title{Quantum Monte Carlo study of the metal to insulator transition on a honeycomb lattice with 1/r interactions}
\author{Li Chen} 
\author{Lucas K. Wagner }
\email{lkwagner@illinois.edu}
\affiliation{Department of Physics, University of Illinois at Urbana-Champaign, Urbana, Illinois 61801, USA.} 
\begin{abstract}

Describing correlated electron systems near phase transitions has been a major challenge in computational condensed-matter physics.   
In this paper, we apply highly accurate fixed node quantum Monte Carlo techniques, which directly work with many body wave functions and simulate electron correlations, to investigate the metal to insulator transition of a correlated hydrogen lattice.
By calculating spin and charge properties, and analyzing the low energy Hilbert space, we identify the transition point and identify order parameters that can be used to detect the transition. 
Our results provide a benchmark for density functional theories seeking to treat correlated electron systems. 
\end{abstract}

\maketitle

\section{Introduction}

Many spectacular phenomena occur near phase transitions of correlated electron systems\cite{dagotto2005complexity}.
For example, high temperature superconductivity\cite{crow1990high, burns1992high, ginzburg1982high}, colossal magnetoresistance \cite{rao1998colossal, ramirez1997colossal} , and the magnetocaloric effect\cite{wada2001giant, pecharsky1997giant, krenke2005inverse} all occur near phase transitions.
An emblematic correlated phase transition is the metal to Mott insulator transition (MIT), which is a metal insulator transition that would not occur in the absence of interactions.
Near this transition, the system is neither in the non-interacting limit, nor in the strongly interacting limit. 

Because there is no small parameter near the MIT, it is challenging to describe the system theoretically.
Single determinant pictures fail qualitatively in this region of physical space\cite{potthoff2012strongly, cohen2011challenges, liechtenstein1995density}.
Exotic states in between the insulator and metal, like the spin liquid state \cite{shimizu2003spin, chandra1988possible,  kalmeyer1989theory}, have been proposed in this region based on approximate theories. 
Whether these states might exist in realistic material systems is still very much an open question because solutions either focus on a very simplified model or make large approximations in the solution of the first principles Hamiltonian. 

Exact correlated solutions can be found for the Hubbard model. 
Sorella and collaborators\cite{sorella2012absence} conducted large scale unbiased quantum Monte Carlo calculation on the honeycomb lattice.
 They showed that there is no evidence for the spin liquid phase near the transition between semi-metal and antiferromagnetic insulator. 
However, this is far from realistic systems since the Hubbard model only includes on-site interactions. 

For the full first principles Hamiltonian, there are no exact solutions. 
There are two broad classes of approaches in this case. The first is density functional theory (DFT) plus corrections, such as LDA+U\cite{liechtenstein1995density, anisimov1997first, anisimov1991band}, and LDA+DMFT\cite{anisimov1997first2, anisimov2005full}.
While these techniques often offer substantial improvement over the underlying DFT calculations\cite{himmetoglu2014hubbard, anisimov1991band, anisimov1997first}, they depend on the starting point, parameter values\cite{held2006realistic}, and have significant uncertainty due to double counting of correlations  \cite{nekrasov2013consistent}.
The second class consists of many-electron wave function techniques, which have no adjustable parameters but are computationally demanding and must approximate the wave function form for efficiency. 
For extended systems, quantum Monte Carlo (QMC) methods, in particular fixed node projector (diffusion or reptation) Monte Carlo is broadly applied, with recent applications\cite{zheng2015computation, busemeyer2016competing} to realistic strongly correlated systems. 
However, the FN-DMC method suffers from the fixed node error, which has not been explored in depth near the metal-insulator transition for realistic periodic systems.

In this study, we investigate the fixed node error of a honeycomb lattice of hydrogen atoms using fixed node reptation Monte Carlo (FN-RMC). 
We choose this system for several reasons. 
First, it is one of the simplest systems with a $1/r$ interaction, and the closest realistic system to a Hubbard model. 
Second, since there is only one electron per atom, we expect that the nodal error will be at its minimum in this system. 
We assess the fixed node error by using nodes from both the metallic and antiferromagnetic insulating mean-field states. 
We investigated five order parameters to identify the transition point: double occupancy, compressibility, staggered moment, spin structure factor and spin spin correlation.  
To find the most accurate ground state quantities, we performed  QMC calculations with multiple starting trial wave functions and find the ground state order parameters by fitting. 
In our data, we could find no evidence of intervening phases; the ground state transitions from a paramagnetic to an antiferromagnetic system at around a lattice constant of $a=2.75$ \AA. 
Our data is appropriate for density functional development, since standard DFT in the PBE functional mispredicts the transition by around 0.2 \AA. 

\section{Method}
First-principle methods start from the Hamiltonian of interacting electrons and ions.
Because electrons and ions do not move on the same time scale, we use the Born-Oppenheimer approximation\cite{born1927quantentheorie} to separate their motion.
The Hamiltonian of many-body electrons system is then
\begin{align}
\hat{H} = - \frac{1}{2m_e}  \sum\limits_i  \nabla_i^2  
          -  \sum\limits_{i, I}  \frac{Z_I e^2}{|\bold{r_i} -\bold{r_\alpha} |}  \notag \\
          +  \frac{1}{2} \sum \limits_{i \neq j}  \frac{e^2}{ | \bold{r_i} - \bold{r_j} |}  
          +  \frac{1}{2} \sum\limits_{I \neq J}  \frac{Z_I Z_J e^2} {\bold{r_\alpha} -  \bold{r_\beta}} , 
\label{eqn:hamiltonian}
\end{align}  
where $i$, $j$ refer to electronic coordinates, and $\alpha$, $\beta$ refer to ionic coordinates. 
This Hamiltonian contains the kinetic energy of electrons, electron-electron interactions, electron-ion interactions and ion-ion interactions.

\subsection{Variational Monte Carlo (VMC)} 

In variational Monte Carlo(VMC), the expectation value of the energy is evaluated by computing the integral
\begin{equation} \label{convex}
E_V(P) = \langle  \Psi_T | \hat{H} | \Psi_T \rangle = \int  dR \frac{{|\Psi_T(P)|}^2}{\int dR {|\Psi_T(P)|}^2}  \frac{\hat{H} \Psi_T(P)} {\Psi_T(P)},
\end{equation} 
where  $\Psi_T$ is the trial wavefunction and $P$ is a list of some parameters.  
Expectation values of observables are calculated by sampling the probability distribution $P(R)= {| \Psi_T |}^2 /  \int dR {|\Psi_T | }^2$ and summing over the sampled values.
We optimize parameters within a VMC trial wavefunction such that the variance of the local energy is minimized.

We constructed compact Slater-Jastrow type trial wavefunctions, which are antisymmetrized products of single-particle orbitals and non-negative Jastrow correlation factors\cite{foulkes2001quantum, kolorenvc2010wave}
\begin{equation} \label{convex}
\begin{array}{cl}
\Psi (R) = e^{J(R, X; P)} D^{\uparrow} (r_{1_\downarrow},..., r_{N_{\downarrow}}) D^{\downarrow} (r_{1_{\uparrow}} , ..., r_{N_\uparrow})
\end{array}
\end{equation}
where $R=(r_1, r_2, ..., r_N )$ are the spatial coordinates of electrons,  $R_\alpha=(r_{\alpha1}, r_{\alpha2},  ..., r_{\alpha M})$ are the spatial coordinates of ions, and $P=({p_1, p_2, ... , p_i})$ are the Jastrow coefficients that must be optimized. 
We generate the Slater determinants with density functional theory \cite{gross2013density, parr1980density, jones1989density}. 
Correlation between electrons is included via the Jastrow factor $J$, which is a two body term, 
\begin{equation} \label{convex}
\begin{array}{cl}
J(R, X; P) = \sum\limits_{i,j}  f(r_i - r_j ; P) + \sum\limits_{i, \alpha} g(r_i- r_\alpha; P). 
\end{array}
\end{equation}
Here $f$ and $g$ refer to electron-electron and electron-ion interactions, respectively.
The Jastrow factor introduces local correlations between electrons that reduce the likelihood they get close to one another and affects the distances of electrons with same spin.

Althrough VMC is easy to implement and computationally efficient, VMC with a single Slater-Jastrow wave function ansatz is not accurate enough. 
As a result, we use the VMC method as precursor to FN-RMC as a means of optimizing trial wavefunctions for later use in more accurate FN-RMC calculations.

\subsection{Fixed Node Reptation Monte Carlo} 
In the diffusion Monte Carlo method, operators that do not commute with Hamiltonian suffer from the mixed estimator error, which is linear in the trial wave function error, 
\begin{align}
\langle A(R) \rangle  & = \frac{\langle \Psi_0 | A(R) | \Psi_0\rangle}{\langle \Psi_0 | \Psi_0 \rangle} \notag \\
 &= \frac{\langle \Psi_T | A(R) | \Psi_0\rangle}{\langle \Psi_T | \Psi_0 \rangle} + \mathcal{O}(|\Psi_T-\Psi_0|). 
\end{align}
A combination of mixed and variational estimators, termed the extrapolated error, reduces the error to second order, 
\begin{align}
\langle A(R) \rangle  & = 2\frac{\langle \Psi_T | A(R) | \Psi_0\rangle}{\langle \Psi_T | \Psi_0 \rangle} 
 - \frac{\langle \Psi_T | A(R) | \Psi_T\rangle}{\langle \Psi_T | \Psi_T \rangle}
 + \mathcal{O}(|\Psi_T-\Psi_0|^2)  \notag \\ 
 &= 2A(R)_{DMC} - A(R)_{VMC} + \mathcal{O}(|\Psi_T-\Psi_0|^2). 
 \label{eqn:mix_extra}
\end{align}

The reptation quantum Monte Carlo (RMC) \cite{pierleoni2005computational, baroni1999reptation} method is a stochastic projection approach that determines the ground state by repeatedly applying the projection operator to a trial wavefunction. 
Comparing with DMC, RMC results are free from mixed estimator error and population control bias. 
The expectation value of a local observable $A(R)$ is calculated as 
\begin{align}
\lim\limits_{\tau \rightarrow \infty} & \frac{\langle  \Psi_T |e^{-\frac{\tau\hat{H}}{2}} A(R_{p/2})   e^{-\frac{\tau\hat{H}}{2}}|\Psi_T\rangle}{\langle e^{-\frac{\tau\hat{H}}{2}} \Psi_T | e^{-\frac{\tau\hat{H}}{2}} \Psi_T \rangle}  \notag \\
&= \frac{\langle \Psi_0 | A(R_{p/2}) | \Psi_0\rangle}{\langle \Psi_0 | \Psi_0 \rangle}, 
\end{align}

The absence of mixed estimator error ensures that results are accurate even for order parameters that do not commute with the Hamiltonian.  
However, RMC suffers from the fermion sign problem. 
We address this by using the fixed-node approximation\cite{cances2006quantum, reynolds1982fixed}, which fixes the nodal surfaces of a wave function during the projection process\cite{wagner2007energetics}. 
In fixed-node QMC,  the accuracy of a calculation depends on the nodal surfaces of the trial wave function and gives an upper bound to the ground state energy.  
Since the Jastrow term is positive, the accuracy of our FN-QMC calculations is determined by the nodal surface of the associated Slater determinant. 
In this paper, we will vary the Slater determinant to minimize the total fixed node energy.

\subsection{Order parameters}
Several order parameters are investigated to identify the MIT transition point and transition order. 
\begin{table}[ht]
\centering 
\caption{Order parameters of the unpolarized UNP to  N\'eel transition}
\resizebox{0.5\textwidth}{!}{
\begin{tabular}{c c}
\hline\hline 
 {Order parameters}  &  Definition \\
\hline 
Local compressibility     &   $ \langle (n_i- \langle n_i \rangle)^2 \rangle $ \\
Double occupancy   &   $\langle  n_{i\uparrow} n_{i \downarrow} \rangle $  \\
Staggered moment   &   $ \langle (S_i - S_j)^2 \rangle $    \\
Spin spin correlation   &  $C_s (L_{max}) =  \frac{1}{N N_{\vec{\tau}_{max}}}  \sum\limits_{R, \vec{\tau}_{max}}  \langle S_R \cdot S_{R + \vec{\tau}_{max}} \rangle $ \\
Spin Structure factor   &       $  S_{AF}  = \frac{1}{N} \langle [ \sum\limits_r  (S_{r, A} - S_{r, B}) ]^2 \rangle $  \\
\hline
\end{tabular}}

\begin{tablenotes}
      \small
      \item  Here $n_i= n_{i\uparrow} + n_{i\downarrow}$ is  the electron density on $i$-th site. $S_i = n_{i\uparrow} - n_{i\downarrow} $ is the spin density on the $i$-th site. In $S_{r, A}$  and  $S_{r, B}$ , $r$ refers to the $r$-th unit cell, A and B indicate different sublattices. 
    \end{tablenotes}
\label{table:Order}
\end{table}

\textbf {\underline{Compressibility:}}  
Compressibility measures the averaged local spin fluctuation on each site. It is defined as
\begin{align}
\langle (n_i- \langle n_i \rangle)^2 \rangle =  \langle( n_{i \uparrow}  + n_{i \downarrow} 
 - \langle n_{i \uparrow} + n_{ i \downarrow}  \rangle )^2\rangle , 
\end{align}
where, $n_i$ , $n_{i\uparrow}$  and $n_{i\downarrow}$ are the number of total electrons, the number of spin-up electrons, and the number of spin-down electrons on the $i$-th site respectively.  
Electrons in the unpolarized UNP state have more freedom than those in the N\'eel state, producing a larger local compressibility for the unpolarized state.

\textbf{\underline{Double occupancy:}} 
Double occupancy evaluates the probability of two opposite spins occupying one site, 
\begin{align}
D = \langle n_{i\uparrow} n_{i \downarrow} \rangle.
\end{align}
We expect the double occupancy decrease with the transformation from spin unpolarized to N\'eel state.

\textbf{\underline{Staggered moment:}} 
The staggered moment is the averaged spin difference between nearest neighbors,  
\begin{align}
\langle (S_i - S_j)^2 \rangle =  \langle (( n_{i \uparrow}  - n_{i \downarrow}) - (n_{j\uparrow} - n_{j \downarrow} ))^2 \rangle ,  
\end{align}
where $i$ and $j$ indicate nearest neighbors.
Because spins are uniformly distributed in the unpolarized state and symmetry-broken in the N\'eel state, we expect the staggered moment to increase with the lattice constant.

\textbf{\underline{Spin-spin correlation at maximum distance:}} 

The spin-spin correlation examines the long range correlation between two symmetry-equivalent sites. The spin-spin correlation order parameter is defined as
\begin{align}
C_s (L_{max}) =  \frac{1}{N N_{\vec{\tau}_{max}}}  \sum\limits_{R, \vec{\tau}_{max}}  \langle S_R \cdot S_{R + \vec{\tau}_{max}} \rangle , 
\end{align}
where $S_R$ is the spin operator at site $R$, and  $\vec{\tau}_{max} $ is a vector that connects two symmetry-equivalent sites with maximum distance in the finite cell.  $N_{\vec{\tau}_{max}}$  is the number of  $\vec{\tau}_{max} $ vectors.   

\textbf{\underline{Spin structure factor:}} 
The spin structure factor also evaluates long range interactions,  
\begin{align}
S_{AF}  = \frac{1}{N} \langle [ \sum\limits_r  (S_{r, A} - S_{r, B}) ]^2 \rangle. 
\end{align}
Here  $S_{r, A} $ and $S_{r, B}$ are spin operators on the $A$ and $B$ sublattices of unit cell $r$.

\section{Calculation Setup}

Our calculation was done in three steps. First, we generated Slater determinants with density functional theory. 
We then multiplied a Jastrow factor to each Slater determinant and optimized the resulting trial wave functions using the VMC method. 
Finally, we used these optimized trial wave functions to perform reptation Monte Carlo energy calculations.
To reduce the fixed-node error, we generate multiple Slater determinants by varying spin states and exchange correlation functionals. 
Density functional theory(DFT) calculations were carried out with the CRYSTAL software suite\cite{dovesi2014crystal14, dovesi2014crystal14manual}.  
QMC calculations were performed with the open source package QWalk \cite{wagner2009qwalk}, using a constant time-step of $0.02$ Hartree$^{-1}$ throughout the RMC projection procedure. We checked smaller timesteps with no change in results.
We sampled lattice constants between 2.4 \AA\ and 3.3 \AA, with a step size of 0.05 \AA.

\begin{figure}
\includegraphics[width=0.5\textwidth]{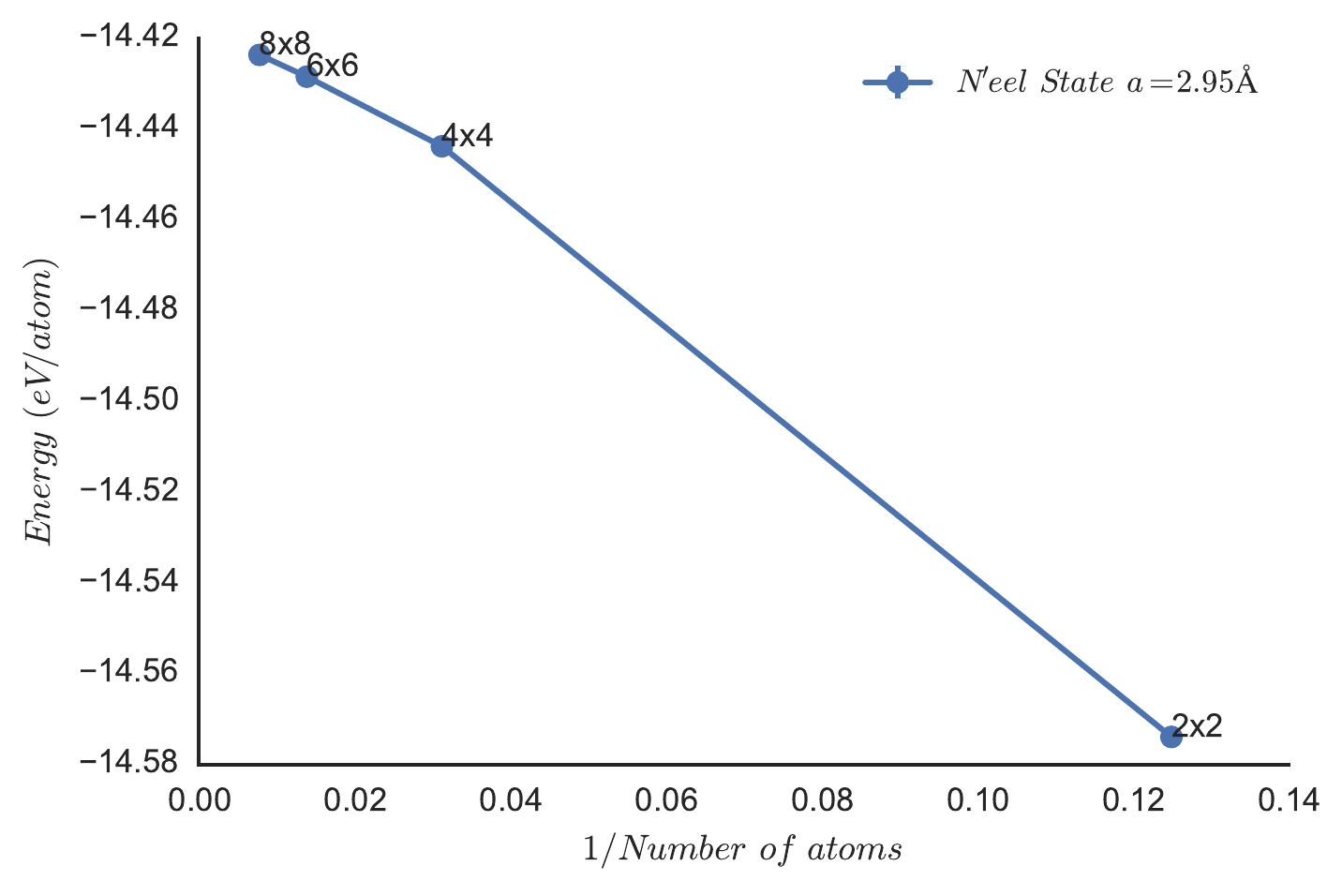}
\caption {Energy vs. 1/number of atoms. Energy starts to converge linearly at lattice cell size 4x4 (32 atoms). }
\label{fig:finite} 
\end{figure}

To control the finite size error, we varied the system cell size (2x2, 4x4, 6x6 and 8x8).
Fig.\ref{fig:finite} shows the influence of finite size error. 
Starting from unit cell containing 32 atoms (cell size 4x4), energy increases linearly with the number of atoms.  
The finite size error for a unit cell with 128 atoms (8x8) has errors in the energy of approximately 1 meV/atom.
Therefore, a unit cell with 128 atoms (8x8) is large enough to reflect the properties of this system.   
In the following section, we report the results for an 8x8 unit cell containing 128 atoms. 

\section{Results and discussion}
\subsection{Trial wave functions from density functional theory} 
\begin{figure}
\includegraphics[width= 0.5\textwidth]{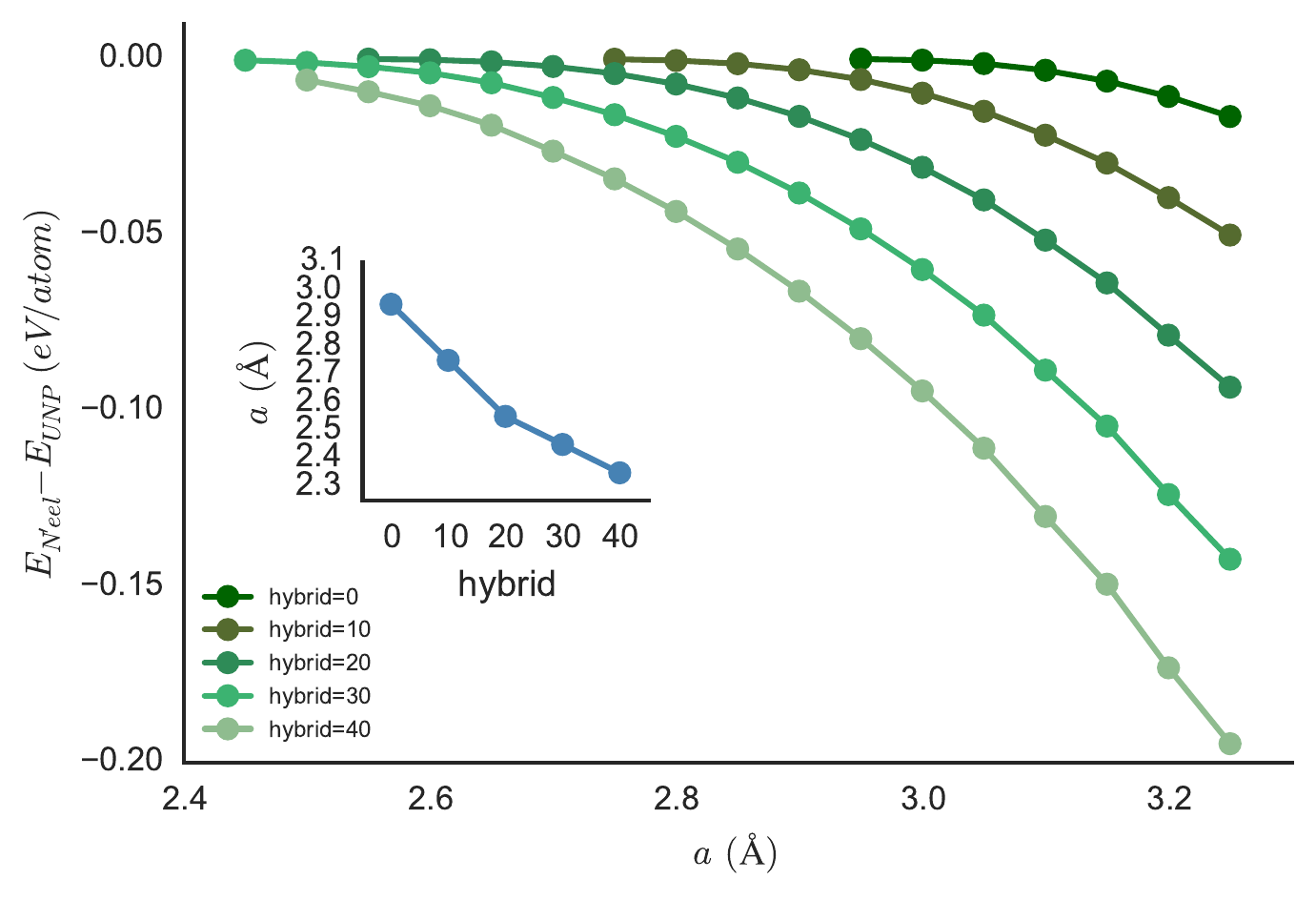}
\caption { DFT energy vs. lattice constant.  Here the vertical axis is the energy difference between N\'eel state and spin unpolarized state.  Lines correspond to different hybridization. The inset plot shows the symmetry breaking point (where the AFM functional produces the N\'eel state) as a function of  hybridization.}
\label{fig:DFTenergy} 
\end{figure}

For small lattice constants, the system is well approximated by a noninteracting model, in which there is no formation of spin moments on the hydrogen atoms. 
Thus one would expect a high quality trial function to be a single Slater determinant with no spin polarization, which we generate using the restricted Kohn-Sham technique. 
We will label this trial wave function UNP, for unpolarized. 
On the other hand, for large lattice constants the system becomes an antiferromagnetic Mott insulator with N\'eel order. 
An appropriate trial wave function for this state is the spin-polarized N\'eel state, in which the spin symmetry is broken and the up/down determinants are inequivalent. 
We term this trial wave function the N\'eel state.

Depending on the density functional used, the N\'eel state may not be stable relative to the UNP state. 
In order to obtain both types of trial function, we used hybrid functionals PBE$_x$ \cite{ernzerhof1999assessment,heyd2003hybrid}, where the functional is given by: 
 \begin{align}
E_{xc}=(1-p) E_x^{PBE} + pE_x^{HF} + E_c^{PBE}.
\label{eqn:hybrid}
\end{align}
The results of these calculations are shown in Fig~\ref{fig:DFTenergy}. 
From a mean-field perspective, one would identify the paramagnetic-antiferromagnetic transition at the point that the N\'eel state becomes lower in energy. 
This transition point is very sensitive to the percentage of Hartree-Fock exchange in the density functional, varying by 0.6 {\AA} over a reasonable range of values.

\subsection{RMC results as a function of the trial wave function.}
\begin{figure}
\includegraphics[width= 0.5\textwidth]{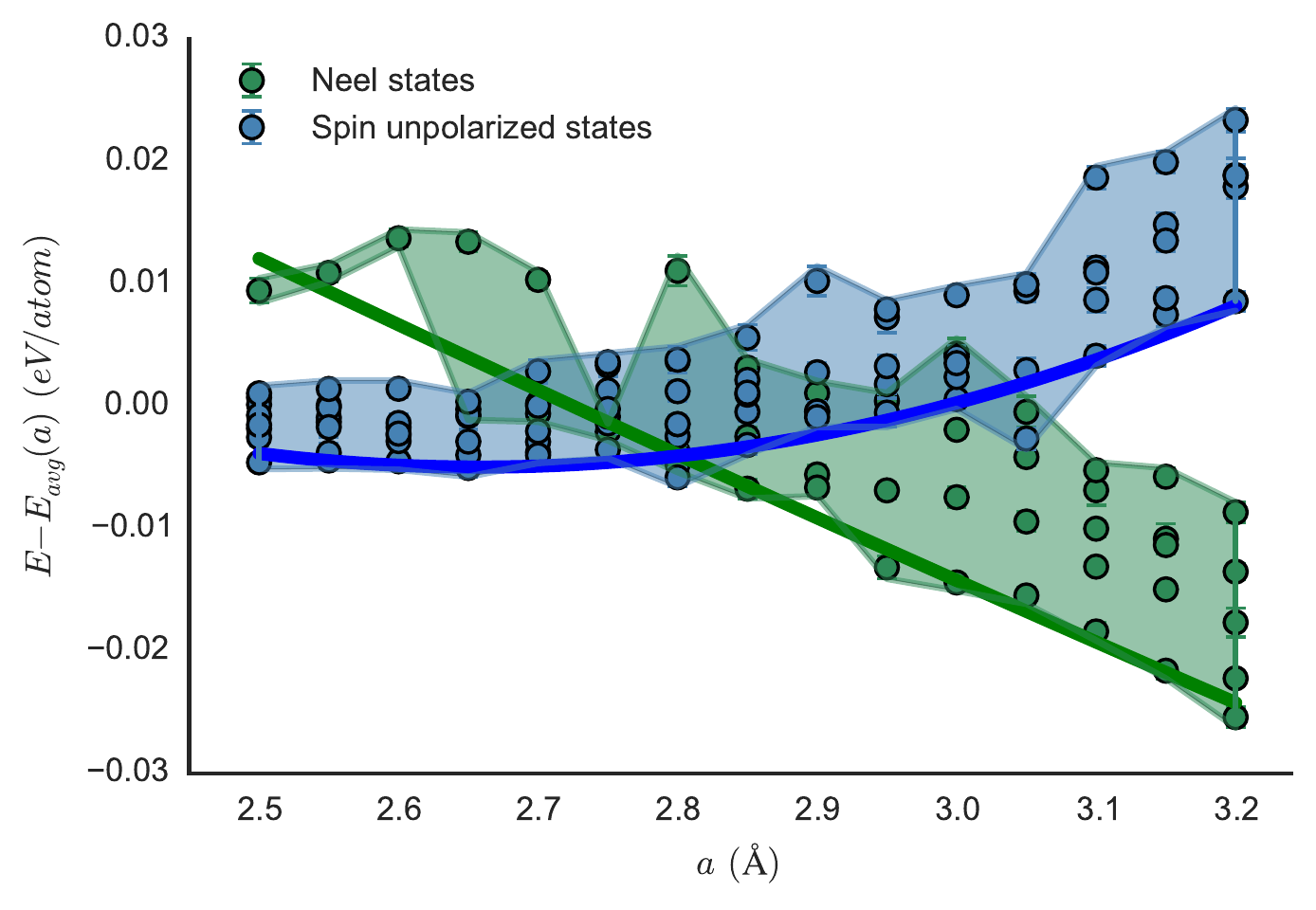}
\caption {Shifted FN-RMC energy vs. lattice constant. Y axis is the shifted energy w.r.t to the averaged value corresponds to the specific lattice constant. 
Blue dots correspond to spin unpolarized states wave functions,  green dots correspond to N\'eel states. 
For clarity, we have drawn regions around trial functions of the same spin state.
 }
\label{fig:RMCenergy} 
\end{figure}

For each value of the lattice constant, we thus have generated a set of Slater determinants that either have spin moments (N\'eel) or are paramagnetic (UNP). 
Fig~\ref{fig:RMCenergy} shows the RMC energy vs. lattice constant for all of these trial functions.
The RMC energies vary by few meV/atom depending on the orbitals.
We mark the lowest energy state of a given type (UNP or N\'eel) by a line on the graph. 
We attempted to use superpositions of UNP and N\'eel states as trial functions, but found no improvement in the energy.

Na\"iively, one might think to determine the paramagnetic-antiferromagnetic transition at the point where the fixed node energy of the minimum of each of the two different trial functions crosses; in this case at around 2.8 \AA. 
However, there are two issues with this approach. 
First, the properties of the fixed node wave function are not guaranteed to be the same as the trial function. 
We have noted several cases, for example VO$_2$ and FeSe\cite{zheng2015computation, busemeyer2016competing}, where a trial function from an insulating mean-field solution results in a zero gap in fixed node diffusion Monte Carlo. 
Second, there is substantial variation of the fixed node energy even within the same class of trial function, which leads to uncertainty in the transition point.

\subsection{Differences between the order parameters of the trial function and the FN-RMC result} 
\begin{figure*}
\includegraphics[width=\textwidth]{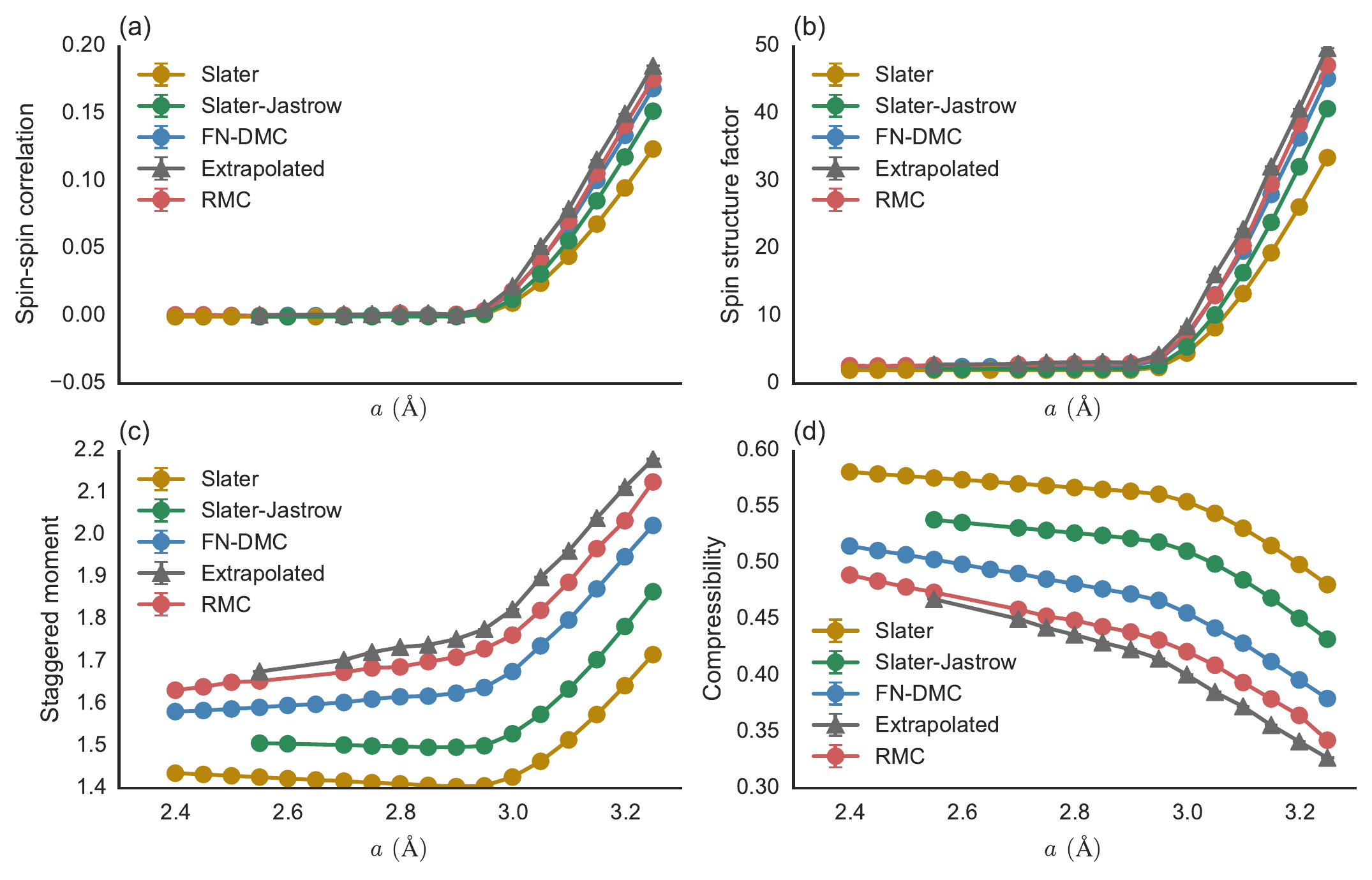}
\caption {Order parameters computed with a PBE trial function.  
Green, blue and red colors represent the results calculated with VMC, DMC and RMC respectively. 
Grey dots are extrapolated values with Equation~\ref{eqn:mix_extra}. 
All statistical uncertainties are much smaller than the symbols.
}
\label{fig:compareVDR} 
\end{figure*}

To investigate the effect of the projection on the wave function, we evaluated VMC, DMC, and RMC calculations using a trial function made up of orbitals from the PBE functional and no hybrid mixing. 
The code was allowed to break symmetry to form a N\'eel state, which happens at around 3 \AA, as can be seen in Fig~\ref{fig:DFTenergy}.
There are immediately several things that are interesting to note about these curves presented in Fig~\ref{fig:compareVDR}. 
First is that the local compressibility is decreased for all lattice constants as we move from a Slater determinant to a correlated wave function. 
This is due to a decrease in double occupancy through short-range correlations. 
Concurrently with this change, the staggered moment increases, since opposite spin electrons spend more of their time on separate sites, even in the metallic phase. 
The long-range order parameters, spin-spin correlation, and spin structure factor, also increase.

At the transition, the Slater determinant has a sharp change in all order parameters. 
As the treatment of correlation improves, the transition becomes more smooth, to the point that it is very difficult to resolve in the local compressibility. 
Given that the orbitals from PBE are not optimal, we can see that the transition point identified using this trial function would be somewhat larger than the optimized wave function presented later in Section~\ref{section:order}, but also somewhat smaller than PBE itself. 
It thus appears that the projection does correct the trial function in the correct direction, but the fixed node error is large enough to prevent a full relaxation.

\begin{figure*}
\includegraphics[width=\textwidth]{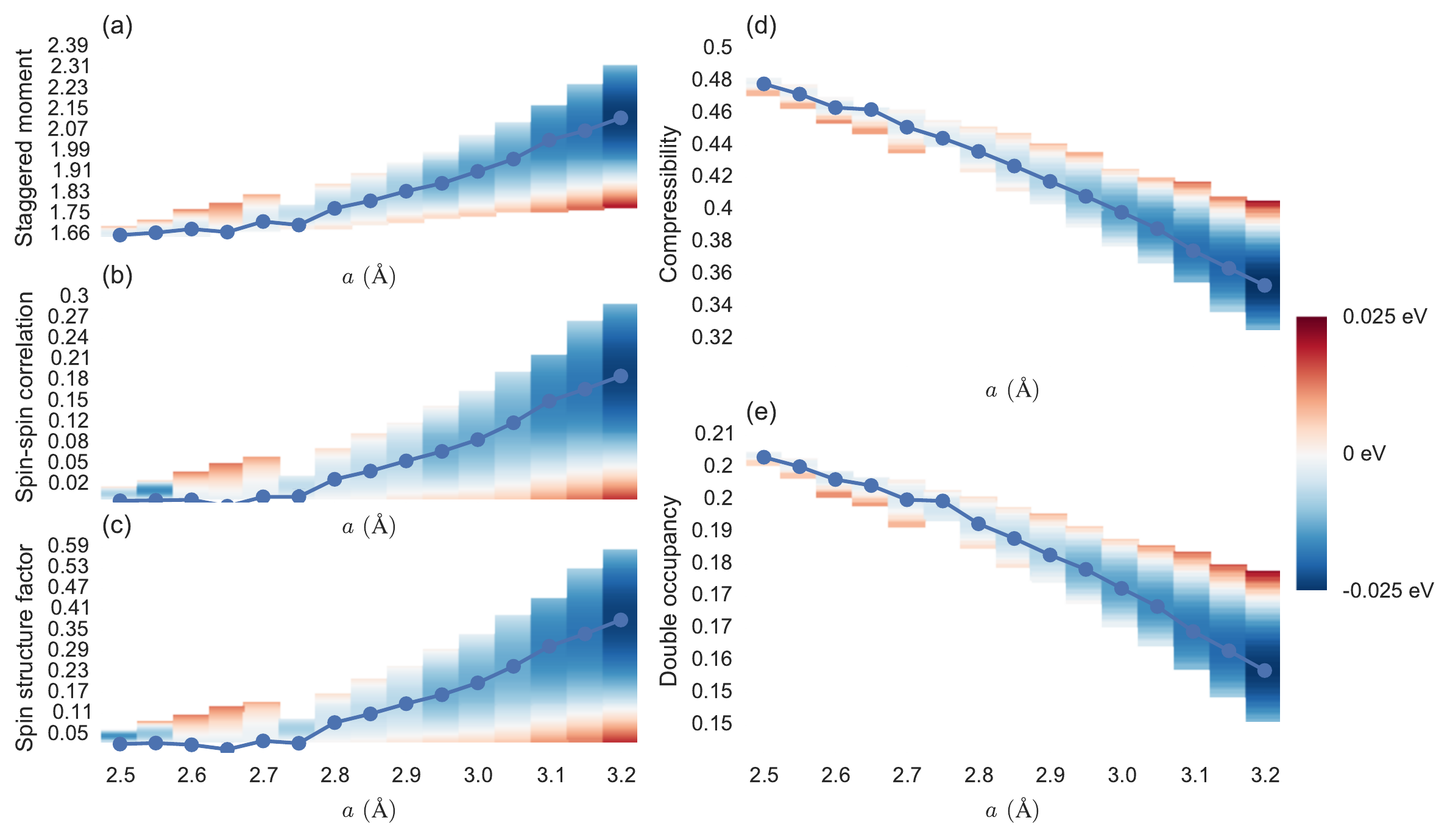}
\caption {Order parameters vs. lattice constant.  Heatmaps are colored by energy; blue represents low energy and red represents high energy. 
The curve on top of the heatmap depicts the fitted minimum energy order parameters.  
}
\label{fig:comparePara} 
\end{figure*}

\subsection{Order parameters} \label{section:order}
\begin{figure*}
\includegraphics[width=\textwidth]{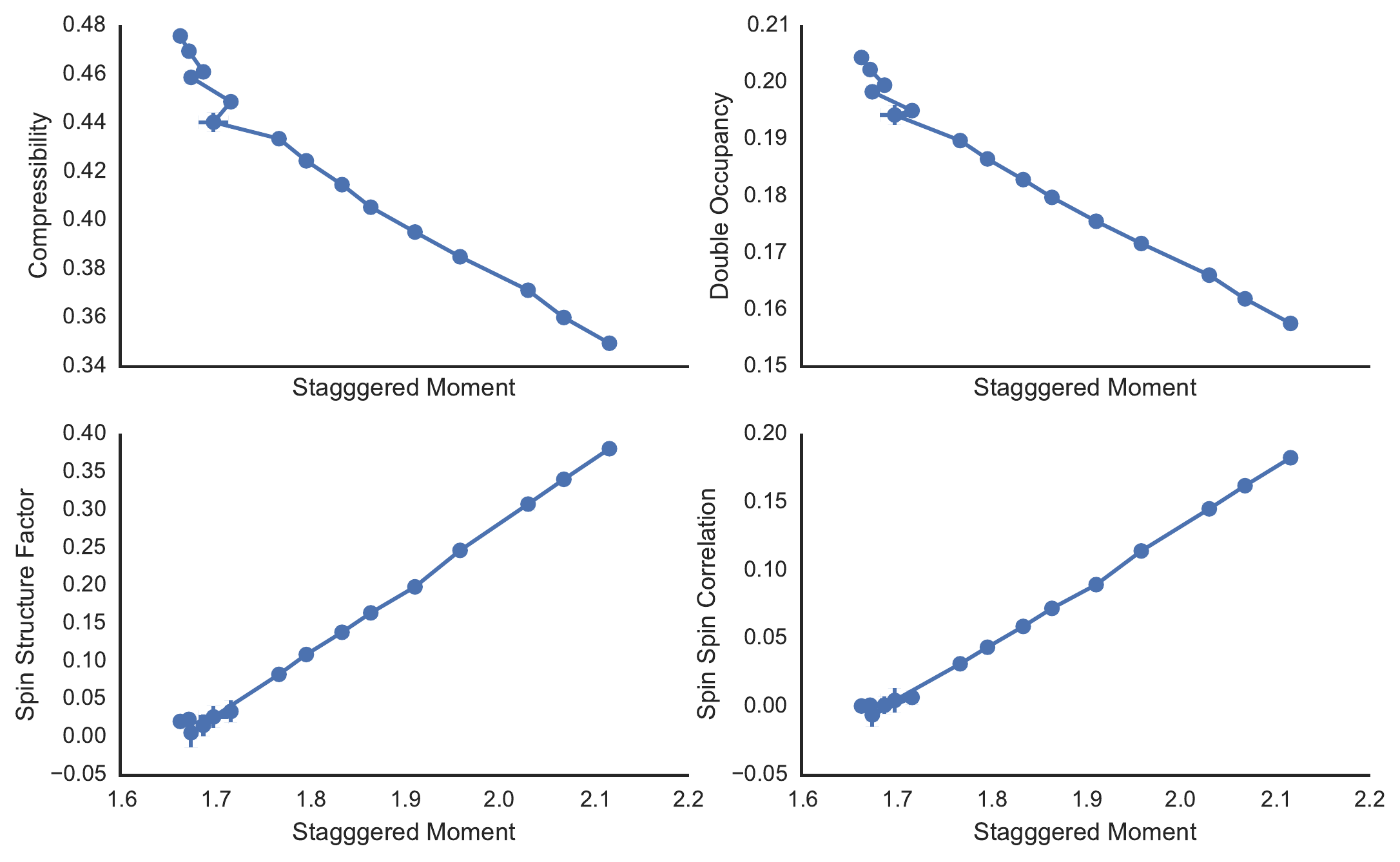}
\caption {  Other order parameters vs. staggered moment. 
The minimum energy wave functions have order parameters that are linearly correlated.}
\label{fig:qvq} 
\end{figure*}

Our partial solution to the dependence of the results on the trial function is to compute the energy as a function of order parameters of the correlated wave function. 
The investigated order parameters are listed in Table.[\ref{table:Order}] and summarized in this section.   
\begin{table*}
\centering
\caption{Correlation coefficients between order parameters} 
\begin{tabular}{ c c c c c}
  \hline
   & Staggered moment & Spin structure factor & Spin-spin correlation  & Double occupancy \\ 
   \hline
Local compressibility   &    -0.98781498  &  -0.97847496  &  -0.97817482 & 0.99875347  \\  
Staggered moment    &         &   0.99791757  &  0.99789502 & -0.99360847  \\
Spin structure factor   &    &   &        0.99985429 & -0.98578002 \\
Spin-spin correlation  &  &   &  &  -0.98553147  \\
\hline
\end{tabular} 
\label{table:Corr}
\end{table*} 
The heatmaps in Fig~\ref{fig:comparePara} shows the calculated order parameters as a function of lattice spacing.
We use blue (red) to denote the lower (higher) energy regions.
We fit the energy as a function of the order parameter and minimize the energy function to estimate the ground state order parameters. 
The curve overlaying the heatmap depicts the fitted ground state compressibility. 

The local compressibility (Fig~\ref{fig:comparePara}(d)) and double occupancy (Fig~\ref{fig:comparePara}(e)) curves are smooth, which indicates a continuous transition.
Obvious kinks show up simultaneously around $a\approx2.75\ \AA$ in the plots of staggered moment (Fig~\ref{fig:comparePara}(a)),  spin-spin correlation (Fig~\ref{fig:comparePara}(b))  and  spin structure factor (Fig~\ref{fig:comparePara}(c)). 
This observation reveals a paramagnetic-antiferromagnetic transition at a critical point around $a=2.75\ \AA$.
From Fig~\ref{fig:DFTenergy}, the transition point identified by DFT calculations varies with the change of exchange correlation functional, so it is difficult to accurately estimate the transition point; our QMC results provide a benchmark for the methods like DFT; it appears that in this case a hybrid of around 20-30\% obtains a transition similar to the QMC result. 
As can be seen from Fig~\ref{fig:comparePara}(c), the RMC calculation can miss the transition if sufficiently poor trial wave functions are used.
We found wave functions that are high in fixed node energy, but have very small spin structure factors. 

Fig~\ref{fig:comparePara} can give some hints as to the nature of the metal insulator transition. 
First, the order parameters of the minimum energy wave functions change continuously as we pass through the transition, with no discernible jumps. 
To the limits of our statistical resolution, the energy also appears to have no first order kinks. 
The computed transition thus appears to be second order, or potentially a crossover. 

To check for intervening phases, we also evaluated the correlation coefficients between different order parameters, with the result shown in Table.[\ref{table:Corr}].  
Fig~\ref{fig:qvq}  shows the correlation between the staggered moment and the other order parameters. 
We find that these order parameters are almost perfectly correlated.  
So it appears that our sampling essentially spans only a one dimensional path through Hilbert space.
We never saw a tendency for the RMC process to move outside this path between metal and antiferromagnetic insulator, which might have happened if there are other phases.
While it is possible that there are other intervening phases, the fixed node error would have to be large enough to prevent the RMC process from accessing them.

\section{ Conclusion} 

We have used fixed node reptation Monte Carlo to study a correlated metal-insulator transition on the honeycomb lattice with $1/r$ interactions. 
The fixed node error in this material is on the order of 10 meV/atom, but can affect the computed properties of the fixed node wave function significantly. 
We addressed this by considering an ensemble of wave functions to map out the low-energy Hilbert space as a function of the order parameters. 
This enabled a clear identification of the metal insulator transition point, which seems to be a continuous transition or a crossover.
We have provided our data which can be used as a high quality benchmark for density functional theory development; not just for the energy but also the properties of the wave function.

\section{ Acknowledgments}
This work was supported by NSF Grant No. DMR 1206242, and the Simons Foundation Collaboration on the many-electron problem.
We acknowledge the computer resources from the campus cluster program at UIUC.
The authors would like to thank David Ceperley for helpful and inspiring discussions. 
We also appreciate the helpful suggestions from our group members, particularly Huihuo Zheng, Brian Busemeyer and Kiel Troy Williams. 

\bibliography{mycite}

\end{document}